\documentclass[12pt]{article}
\usepackage{epsfig}
\usepackage{url}

\usepackage{array,tabularx,epsfig,mathrsfs,graphicx,rotating}
\usepackage{ifthen}
\usepackage{fancybox}
\usepackage{amsfonts}

\newcommand{\Ks}{K^{0}_{_{S}}}
\chardef\til=126
\newcommand{\gev}{{\,\mathrm{GeV}}}
\newcommand{\pb}{{\,\mathrm{pb}}}
\date{}

\begin{document}
\title{Measurements of  $K^{\pm},\,\Ks$,  $\Lambda$ and $\bar{\Lambda}$ 
and Bose-Einstein Correlations between Kaons  at ZEUS}

\author{B.B. Levchenko
\thanks{Partly supported by the Russian Foundation for Basic Research,  grant 
no. 05-02-39028-NSFC-a.}
\footnote{The talk presented at  XV Workshop on Deep-Inelastic Scattering and Related Subjects 
(DIS07), Munich, Germany, April  16-20, 2007 and 7-th Workshop  on  Very High Multiplicity Physics
 (VHMP), Dubna, Russia, September 17-19, 2007.}
\\
{(\em On behalf of the ZEUS Collaboration)}
%
\vspace{.3cm}\\
%
Skobeltsyn Institute of Nuclear Physics,  Moscow State University\\
119991 Moscow, Russian Federation
 }
\maketitle
\begin{abstract}
  Measurements of  production of the neutral and charged strange hadrons  in $e^{\pm}p$ 
  collisions  with the ZEUS detector are presented. 
The data on differential cross sections, baryon-to-meson ratios, baryon-antibaryon  asymmetry  
and Bose-Einstein correlations in deep inelastic scattering and photoproduction are 
summarized \cite{url}.
\end{abstract}

\section{Introduction}
After pions, strange hadrons are the most copiously produced particles
in  $e^{\pm}p$ collisions with a centre-of-mass energy of 318 GeV at HERA. 
In phenomenological models based on the Lund string scheme \cite{Andersson}, 
an intensity  of strange quark production 
is regulated by a free parameter $\lambda_{s}$, which has a value in the range from 
0.2 to 0.4 for different processes.

The experimental results on $K^{\pm},\,\Ks$, $\Lambda$, and $\bar{\Lambda}$ production 
\cite{zeus06ks,zeus07bec}
presented in this note are based on a data sample of 121$\pb^{-1}$ collected 
by the ZEUS experiment at HERA.
This is about 100 times larger data sample than used in previous HERA 
publications  \cite{zeus98ks,zeus95ks} and extend  the kinematical region of the measurements, thereby 
providing a tighter  constraint on models.

\section{Measurements of $\Ks$, $\Lambda$ and $\bar{\Lambda}$}
Weak decaying neutral $\Ks$ and  $\Lambda$ are well reconstructed in the
modes $\Ks\rightarrow \pi^{+}\pi^{-}$, $\Lambda\rightarrow p\pi^{-} $, 
$\bar{\Lambda}\rightarrow \bar{p}\pi^{+}$ via displaced secondary vertices.
The measurements have been performed in three different regions of $Q^{2}$:  
deep inelastic scattering (DIS) with $Q^2 > 25 \gev^2$; DIS with 
$5 < Q^2 < 25 \gev^2$; and photoproduction (PHP), $Q^2 \simeq 0 \gev^2$. 
In the PHP sample, two jets, each of at least 
$5 \gev$ transverse energy, were required.

\subsection{Spectra of $\Ks$ and  $\Lambda + \bar{\Lambda}$ in DIS}

\begin{figure}[h]
\begin{minipage}{.75\textwidth}
\begin{center}
\includegraphics[width=0.99\columnwidth]{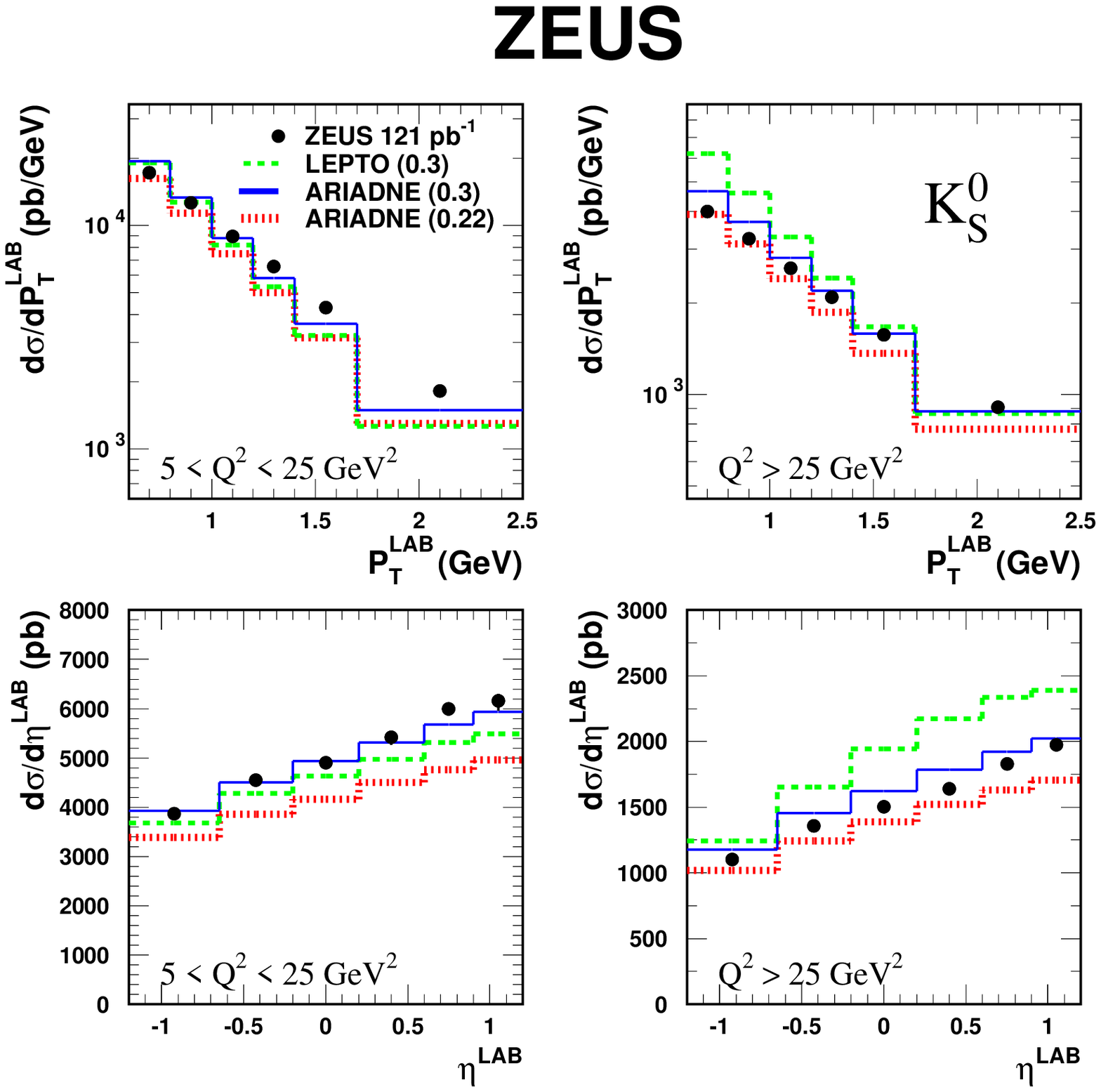}
\includegraphics[width=0.99\columnwidth]{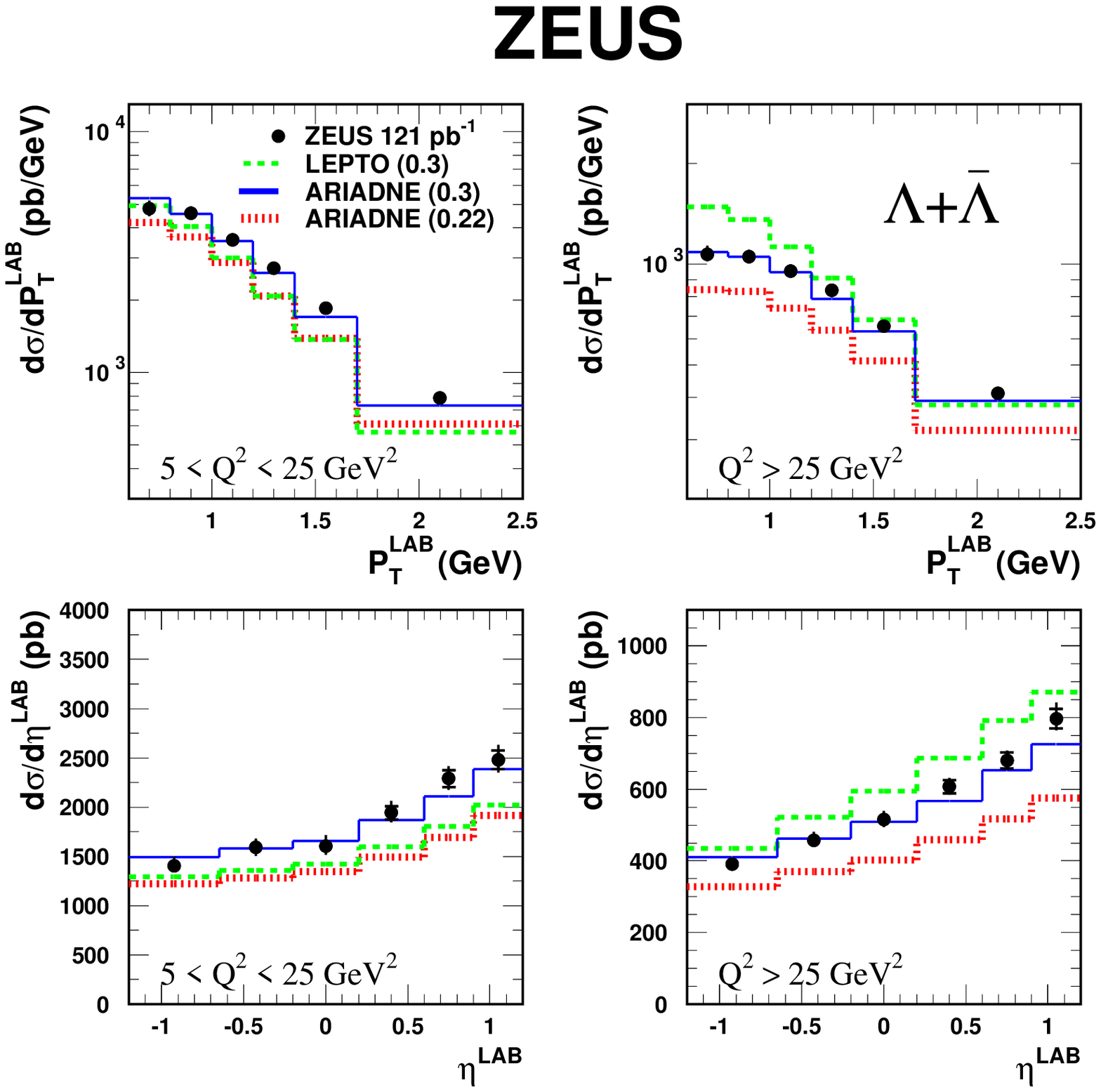}
\end{center}
\end{minipage}
\caption{Differential $\Ks$ and $\Lambda + \bar{\Lambda}$  production cross-sections. 
  The model predictions are  at values of  a strangeness suppression factor
 $\lambda_{s}$ shown in paranthesis.}\label{Fig:1}
\end{figure}

Measured differential cross sections  are shown in Fig.~\ref{Fig:1}. 
 The cross sections are compared to the absolute 
predictions of  {\sc Ariadne 4.12} \cite{MC-A} and {\sc Lepto 6.5} \cite{MC-L} MC calculations. 
The {\sc Ariadne} program with $\lambda_{s}=0.3$ describes the $\Lambda + \bar{\Lambda}$  data 
reasonably well in both $Q^2$ samples. 
The description of the $\Ks$ data  by {\sc Ariadne} is less satisfactory.
The slope of the $P_{T}^\mathrm{LAB}$ dependence  is incorrect and in the high-$Q^2$ domain
the data  requires $\lambda_{s}<0.3$. The cross section at low $x_{_\mathrm{Bj}}$  (not shown) is 
underestimated for both  the low- and the high-$Q^2$ samples \cite{zeus06ks}.
The {\sc Lepto} MC does not describe the data well and predicts a too fast grow of the cross sections 
with $Q^2$.
We conclude, that for the  production of baryons the data reqiuers $\lambda_{s}$ to be approximately constant,
but in  case of $\Ks$ production  $\lambda_{s}$ has to decrease with $Q^2$.

\subsection{Baryon-antibaryon asymmetry in DIS and PHP}

A positive asymmetry of 3.5\% is predicted in DIS \cite{bzk}, due to the so-called gluon-junction mechanism 
that makes it possible for the {\em baryon number to travel} several units of rapidity, in this case from the proton 
beam direction to the rapidity around 0 in the laboratory frame. 

The baryon-antibaryon asymmetry 
$$\mathcal{A}=\frac{N(\Lambda)-N(\bar{\Lambda})}{N(\Lambda)+N(\bar{\Lambda})}$$ 
has been measured and compared 
to MC predictions from {\sc Ariadne}, {\sc Lepto} and {\sc Pythia} \cite{MC-P}. The following value was
obtained at high $Q^2$ : $\mathcal{A} = 0.3 \pm 1.3^{+0.5}_{-0.8} $\%  which has to be compared to 
the {\sc Ariadne} ($\lambda_{s} = 0.3$) prediction of $0.4 \pm 0.2 $\%.
In PHP, $\mathcal{A}$ = $-0.07 \pm 0.6^{+1.0}_{-1.0} $\%, compared to
the {\sc Pythia} prediction of $0.6 \pm 0.1 $\%.

\begin{figure}[h]
\begin{minipage}{.75\textwidth}
\begin{center}
\includegraphics[width=0.99\columnwidth]{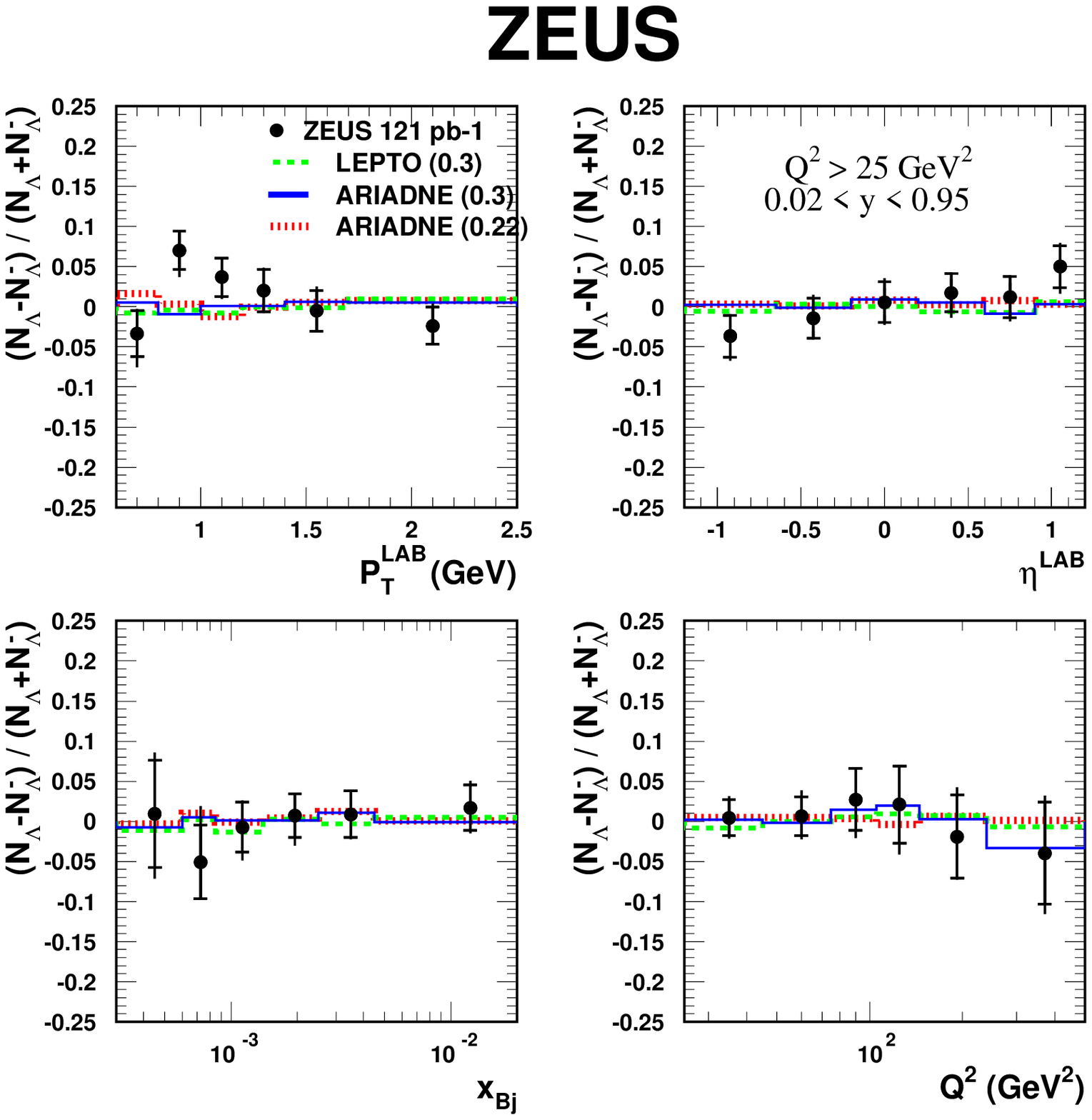}
\includegraphics[width=0.99\columnwidth]{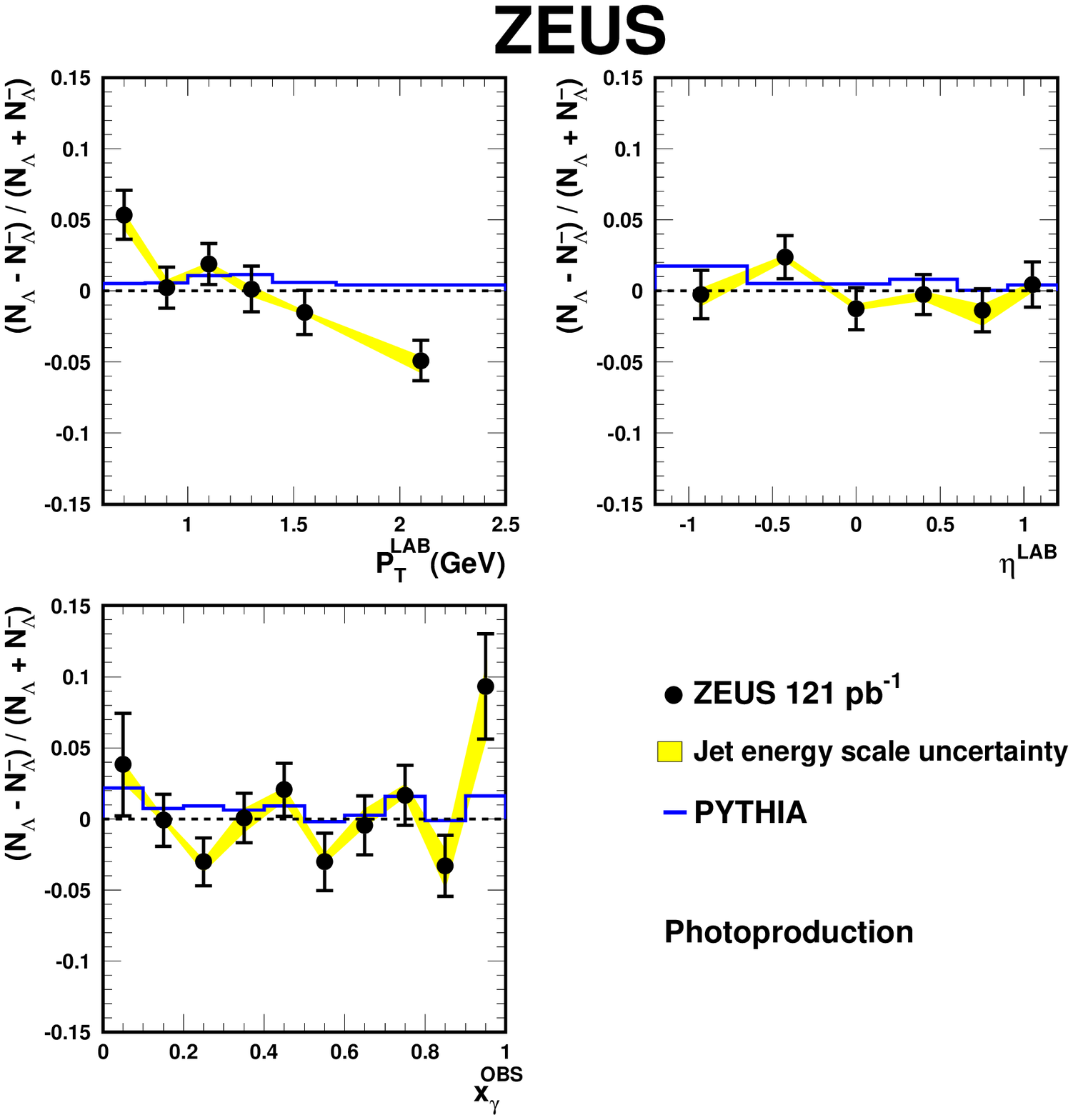}
\end{center}
\end{minipage}
\vspace*{-2mm}
\caption{ The baryon-antibaryon asymmetry $\mathcal{A}$ as a function of $P_{T}^{\mathrm{LAB}}$,
 $\eta^{\mathrm{LAB}}$,
$x_{_\mathrm{Bj}}$ and $Q^2$ for the DIS sample (top),
 and  as a function of  $P_{T}^{\mathrm{LAB}}$,  $\eta^{\mathrm{LAB}}$ and  $x_{\gamma}^\mathrm{OBS}$ for the
photoproduction sample   (bottom). The different lines are the predictions of the different MC  generators,
as indicated in the plots. }\label{Fig:2}
\end{figure}

Figure \ref{Fig:2}  shows   $\mathcal{A}$  at high-$Q^2$ and in PHP.
In all cases, $<\!\mathcal{A}\!>$ is consistent 
both with no asymmetry and  with a very small asymmetry predicted by the Monte Carlo. 
However, as shown in Figs~\ref{Fig:2}, in DIS  the baryon-antibaryon asymmetry
becomes positive and increases in the incoming proton  hemisphere ($\eta^{LAB} >0$),  
 as well as at  $P_{T}^{LAB}$  below 1$\gev$.

\subsection{Baryon-to-meson ratio in photoproduction}

\begin{figure}[h]
\begin{minipage}{.75\textwidth}
\begin{center}
\includegraphics[width=0.99\columnwidth]{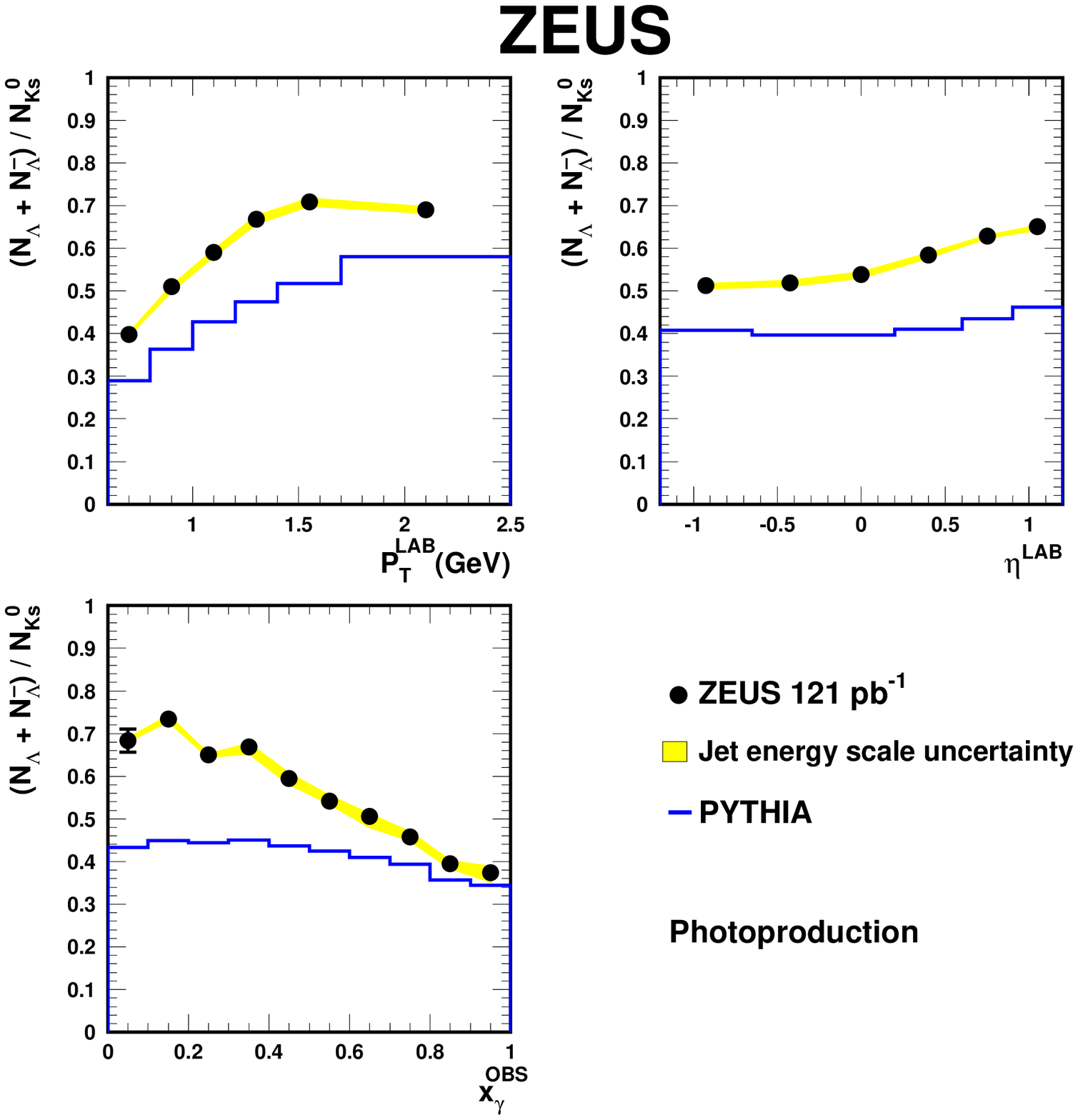}
\includegraphics[width=0.99\columnwidth]{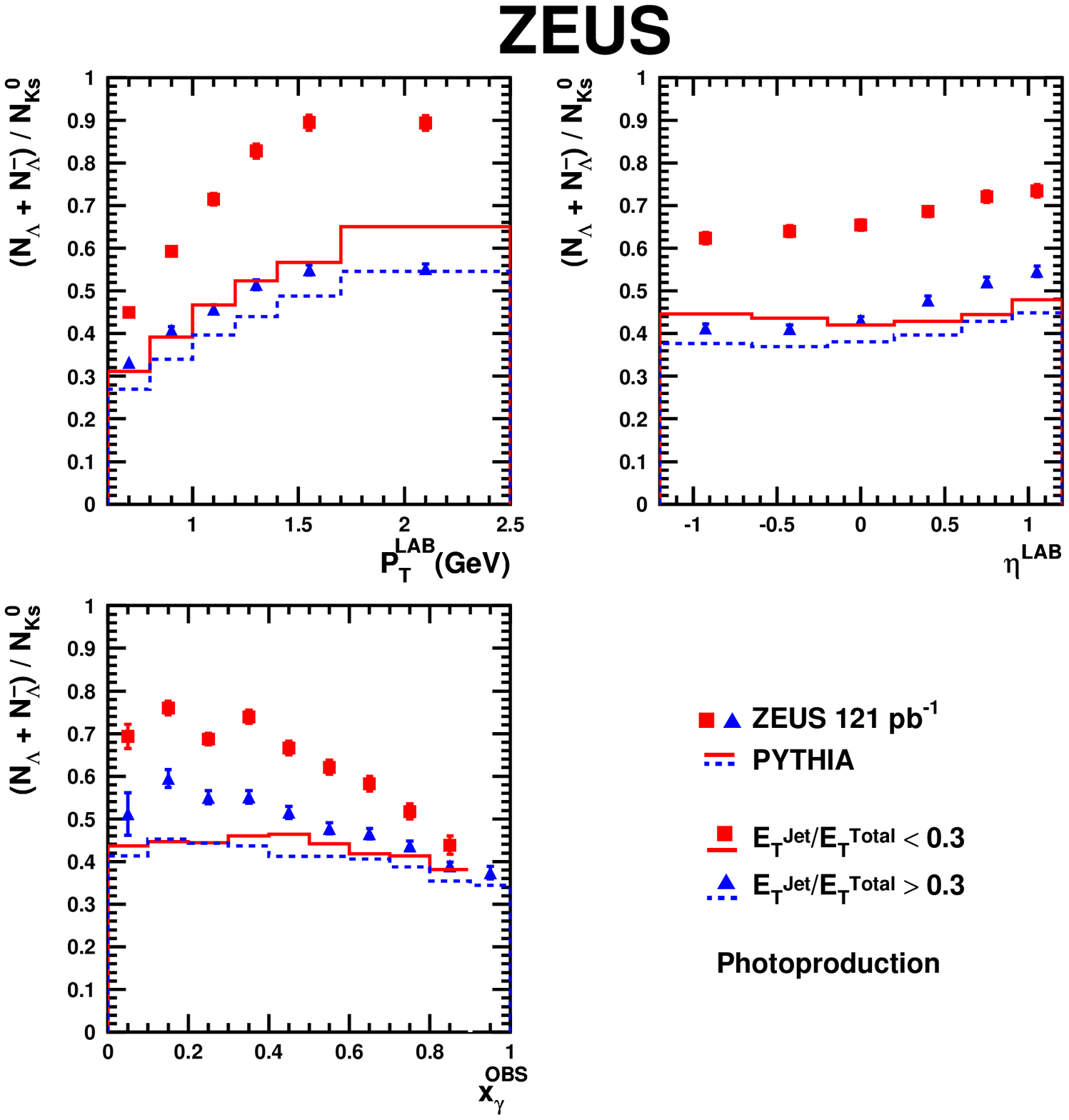}
\end{center}
\end{minipage}
\caption{The ratio $\mathcal{R}$ as a function of $P_{T}^{LAB}$, $\eta^{LAB}$,
and $x_{\gamma}^\mathrm{OBS}$ for the PHP events.  
Top: the ratio from the whole PHP sample.
Bottom: the ratio from the fireball-enriched (squares) and the fireball-depleted 
(triangles) samples. The predictions from {\sc Pythia} for $\lambda_{s}=0.3$.
}\label{Fig:3}
\end{figure}

The relative yield of strange baryons and mesons was studied with the  ratio 
$$\mathcal{R}= \frac{N(\Lambda) + N(\bar{\Lambda})}{N(\Ks)}.$$
Figure \ref{Fig:3} shows $\mathcal{R}$ for the PHP sample. For the direct-enriched sample, 
where  $x_{\gamma}^\mathrm{OBS} >$~0.75, $\mathcal{R}$ is about 0.4, the same value as 
in DIS at low $x_\mathrm{Bj}$ and low $Q^{2}$ \cite{zeus06ks}.  
However, $\mathcal{R}$ rises to a value of about 0.7 towards low $x_{\gamma}^\mathrm{OBS}$ 
(resolved-enriched sample), while it stays flat in the {\sc Pythia} prediction.

In order to study this effect further, the PHP events were divided into 
two samples . In the first, called {\em fireball-enriched},
the jet with the highest transverse energy was required to contribute at most 30\% to 
the total hadronic transverse energy. The other sample, containing all the other events, 
was called {\em fireball-depleted}. 
The measured $\mathcal{R}$  (see Fig. \ref{Fig:3}, Bottom)
is larger for the fireball-enriched sample, most significantly at high $P_{T}^\mathrm{LAB}$, than it is 
for the fireball-depleted sample. This feature is not reproduced by {\sc Pythia}, which predicts almost 
the same $\mathcal{R}$ for both samples. The {\sc Pythia} prediction reasonably describes the measured 
values of $\mathcal{R}$ for the fireball-depleted sample. 
This is not surprising as {\sc Pythia} generates jets in events according 
to the multiple interaction mechanism, 
which makes several independent jets, like those in DIS or $e^+ e^-$ 
where baryons and mesons are created locally. 

\vspace*{3mm}
\hspace*{3mm} We note  that  the increase of the ratio $\mathcal{R}$  
toward the proton  hemisphere, reflects  a  rapid growth  
of the $\Lambda + \bar{\Lambda}$ cross section   as $\eta^{LAB}$ increases, 
as compared to the $\Ks$ cross section grow \cite{zeus06ks}.

\section{Bose-Einstein correlations of charged and neutral kaons in DIS }

\begin{figure}
\begin{center}
\includegraphics[width=0.75\columnwidth]{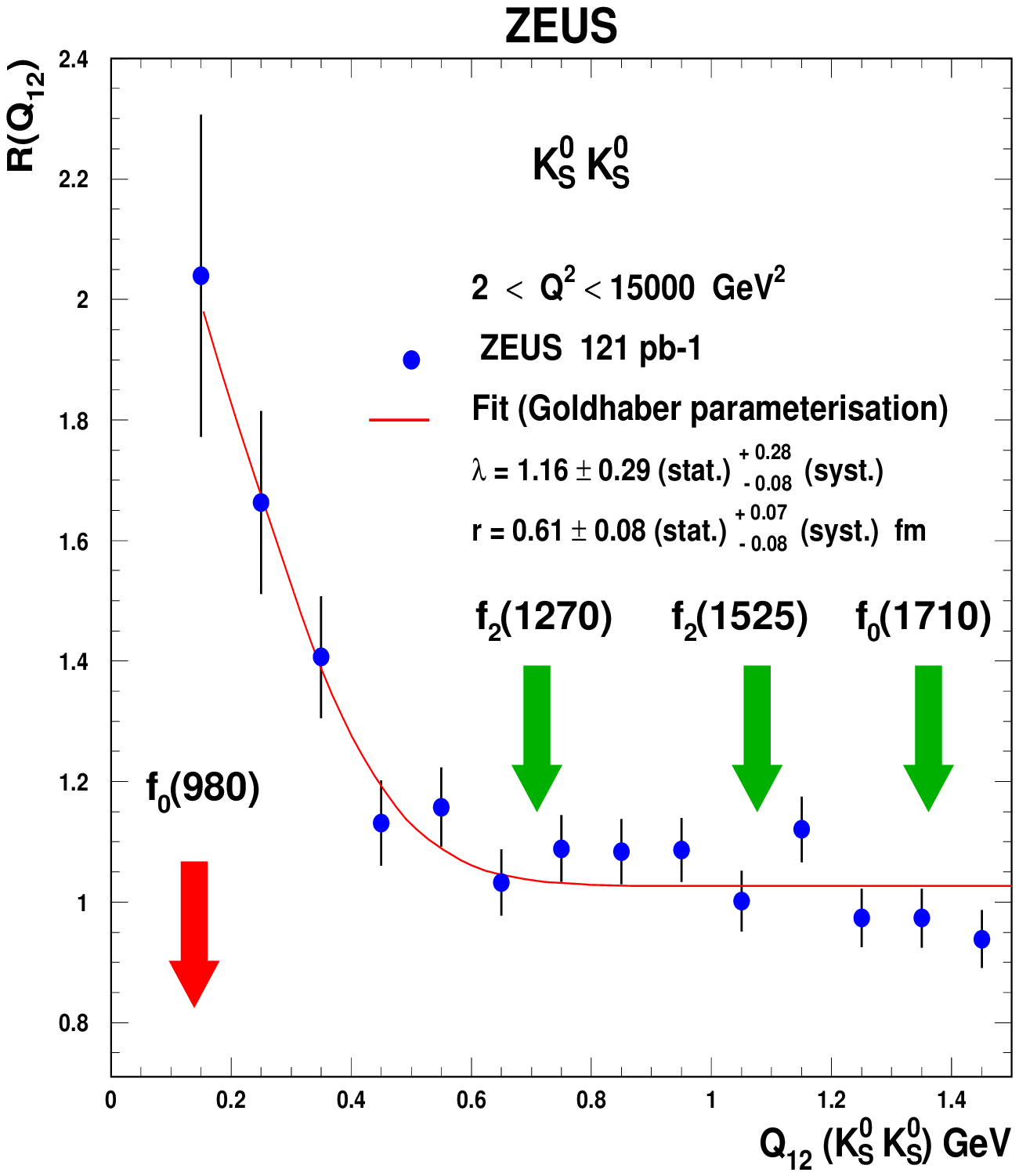}
\caption{ The two-particle correlation functions at $<\!\!Q^2\!\!>=35\gev^2$ for  
 neutral kaons with fits to the Goldhaber function. Arrows indicate $Q_{12}$
regions with contributions from resonances in the  $\Ks\Ks$  system.
  }\label{Fig:4}
\end{center}
\end{figure}

Primordial quantum correlations between identical bosons, so-called Bose-Einstein
correlations (BEC), so far is the only method to estimate the space-time geometry of
an elementary particle emission source. The measurements of the radius of the emission 
source have been mostly performed with pure quantum states 
$\pi^{\pm},\,K^{\pm},\,p/\bar{p}$.~For 
mixed quantum states, like $\Ks$, the information is scare.

The results presented below were obtained  with charged kaons  
selected using the energy-loss measurements, $dE/dx$. The identification
of $K^{\pm}$ is possible for $p<0.9\gev$. The resulting data sample contained 55522 
$K^{\pm}K^{\pm}$ pairs. The $\Ks$ mesons were identified via displaced secondary
vertices. After all cuts, the selected data sample contained 18405 $\Ks\Ks$
pairs and 364 triples \cite{zeus07bec}.

Figure~\ref{Fig:4} shows the two-particle correlation function $R(Q_{12})$ for  identical 
kaons  calculated using the double ratio method 
$$R(Q_{12})=\frac{R_{data}(Q_{12})}{R_{MC}(Q_{12})},$$ 
where $R_{data}(Q_{12})$ is the ratio of the two-particle densities constructed  from pairs 
of kaons coming from the same  and different events. 
$R_{MC}(Q_{12})$ is obtained in a similar way for {\sc Ariadne} MC events without BEC. 
$Q_{12}$ is given by $ Q_{12}=\sqrt{-(p_{1}-p_{2})^2}$. Assuming a Gaussian
shape of emission source, $R(Q_{12})$  were fitted by the standard Goldhaber-like
function 
$$R(Q_{12})=\alpha (1+\lambda\, e^{-Q_{_{12}}^2 r^2})$$
to extract the degree of 
the source coherence $\lambda$ and the source radius $r$. The measured radii  for 
$K^{\pm}K^{\pm}$ and $\Ks\Ks$ are close to each other \cite{zeus07bec}.
 In case of $\Ks\Ks$, the fit (see Fig.~\ref{Fig:4}) does not
take into account a possible contamination from the scalar $f_{0}(980)$  
decaying below the threshold. 
The most  probable fraction of $f_{0}(980)$ which allows to describe the excess of data over MC 
was estimated to be $4\%$. The results corrected for the $f_{0}$ contamination are
$\lambda=0.70{\pm}0.19^{+0.47}_{-0.53}$ and $r=0.63{\pm}0.09^{+0.11}_{-0.08}\,$fm.
Thus, the $f_{0}(980)\rightarrow \Ks\Ks$ decay can significantly affect the $\lambda$ parameter
for $\Ks\Ks$ correlations. The radius-values obtained in DIS agree with $e^+e^-$ annihilation 
results at LEP~\cite{zeus07bec}.
\vspace*{2mm}

{\bf Acknowledgments.} The author is grateful to H. Abramowicz and A. Savin
for reading of the manuscript and comments.

\newpage
 

\begin{footnotesize}

%
\end{footnotesize}


\end{document}